\documentclass[structabstract]{aa}
\usepackage{graphicx}
\usepackage{txfonts}
\usepackage{lscape}

\begin{document}

\title{Radio and IR study of the massive star-forming region 
IRAS~16353$-$4636}

\author{P. Benaglia\inst{1,2}, 
        M. Rib\'o\inst{3},
        J. A. Combi\inst{1,2},
        G. E. Romero\inst{1,2},
        S. Chaty\inst{4},
        B. Koribalski\inst{5},
        I. F. Mirabel\inst{4,6},
        L. F. Rodr\'{\i}guez\inst{7},
        G. Bosch\inst{8}
}
\institute{Instituto Argentino de Radioastronom\'{\i}a, C.C.5, (1894) Villa
Elisa, Argentina\\
\email{[paula;jcombi;romero]@iar-conicet.gov.ar}
\and Facultad de Cs. Astron\'omicas y Geof\'{\i}sicas, UNLP, Paseo del Bosque
s/n, (1900) La Plata, Argentina\\
\email{[pbenaglia;romero]@fcaglp.unlp.edu.ar}
\and Departament d'Astronomia i Meteorologia and Institut de Ci\`ences del Cosmos 
(ICC), Universitat de Barcelona (IEEC - UB), 
Mart\'{\i} i Franqu\`es 1, 08028 Barcelona, Spain\\
\email{mribo@am.ub.es}
\and Laboratoire AIM (UMR 7158 CEA/DSM-CNRS-Universit\'e Paris Diderot),
Irfu/Service d'Astrophysique, Centre de Saclay, B\^at. 709 FR-91191 
Gif-sur-Yvette Cedex, France\\
\email{chaty@cea.fr}
\and Australia Telescope National Facility, CSIRO, PO Box 76, Epping, NSW 1710,
Australia\\
\email{Baerbel.Koribalski@csiro.au}
\and Instituto de Astronom\'{\i}a y F\'{\i}sica del Espacio (IAFE), C.C. 67, Suc. 
28, (1428) Buenos Aires, Argentina 
\email{fmirabel@eso.org}
\and Centro de Radioastronom\'{\i}a y Astrof\'{\i}sica, Universidad Nacional
Aut\'onoma de M\'exico, Morelia 58089, M\'exico\\
\email{l.rodriguez@crya.unam.mx}
\and IALP, UNLP-CONICET, Argentina\\
\email{guille@fcaglp.unlp.edu.ar}
}

\authorrunning{Benaglia et~al.}
\titlerunning{The star forming region IRAS 16353$-$4636}

\date{Received / Accepted}

\abstract 
{With the latest infrared surveys, the number of massive protostellar 
  candidates has increased significantly. New studies have posed additional
  questions on important issues about the formation, evolution, and other 
  phenomena related to them. Complementary to infrared data, radio 
  observations are a good tool to study the nature of these objects, and
  to diagnose the formation stage.}
{Here we study the far-infrared source IRAS 16353--4636 with
  the aim of understanding its nature and origin. In particular, we search
  for young stellar objects (YSOs), possible outflow structure, and 
  the presence of non-thermal emission.}
{Using high-resolution, multi-wavelength radio continuum data obtained
  with the Australia Telescope Compact Array\thanks{The Australia 
  Telescope Compact Array is funded by the Commonwealth of Australia 
  for operation as a National Facility by CSIRO.},
  we image IRAS 16353--4636 and its environment from 1.4 to 19.6 GHz, 
  and derive the distribution of the spectral index at maximum angular
  resolution. We also present new $JHK_{\rm s}$ photometry and 
  spectroscopy data obtained at ESO NTT\thanks{Based on observations
  collected at the European Organisation for Astronomical Research in the
  Southern Hemisphere, Chile (ESO Programme 073.D-0339, PI S. Chaty).}. 
  $^{13}$CO and archival \ion{H}{i} line data, and infrared databases 
  (MSX, GLIMPSE, MIPSGal) are also inspected.}
{The radio continuum emission associated with IRAS 16353--4636 
  was found to be extended ($\sim$10 arcsec), with a bow-shaped  
  morphology 
  above 4.8 GHz, and a strong peak persistent at all frequencies.
  The NIR photometry led us to identify ten near-IR sources and
  classify them according to their color. We used the
  \ion{H}{i} line data to derive the source distance, and analyzed 
  the kinematical information from the CO and NIR lines detected.}
{We have identified the source IRAS 16353$-$4636 as a new protostellar 
  cluster. In this cluster we recognized three distinct sources: 
  a low-mass YSO, a high-mass YSOs, and a mildly confined region of 
intense
  and non-thermal radio emission. We propose the latter corresponds to 
  the terminal part of an outflow.}


\maketitle

\section{Introduction} \label{introduction}

The availability, in recent years, of infrared survey products like 
the MSX (Price et al. 2001) and succesive catalogs have provided an 
extremely rich database to look for massive young stellar objects. 
Programs such as the Red MSX Survey (RMS; Lumsden et al. 2002, Hoare et 
al. 2005) have been successful in finding thousands of candidates. 
The goals of these surveys have been to confirm the evolutionary
status of luminous, embedded sources, to perform a statistical
analysis of these objects on galactic scales, and to look for
efficient mechanisms for massive star formation.

Observations of star forming regions and associated outflows at radio
wavelengths can be used to further our understanding of massive star
formation, a topic which is still not completely understood. However,
such studies require high angular resolution and sensitivity, because
massive young stars are often found at kilo-parsec distances and are
usually associated with densely populated clusters of intermediate and
low mass stars.

Up to now, only a few massive young stellar objects (massive YSOs or 
MYSOs) are related to collimated jets mapped at the radio range, 
such as the Serpens sources (Rodr\'{\i}guez et al. 1989), HH 80-81 
(Mart\'{\i} et al. 1993), and IRAS 16547-4247 (Garay et al. 2003). Garay 
et al. found this last object to be a triple quasi-linear radio source, 
that shows non-thermal indices at the lobes. The source could be 
associated with a molecular core. The authors have proposed that it 
is a MYSO ejecting a collimated wind that interacts with the surrounding 
interstellar medium, producing shocks and consequent radio emission. In 
recent years, near-IR and (sub)millimeter line studies have proved very 
valuable in supplying more information on (M)YSOs; see, for example, 
Varricatt et al. (2010) and references therein.

We have found an intense infrared source, \object{IRAS~16353$-$4636} 
($l,b = 337.99^{\circ},0.08^{\circ}$) which seems to share some 
characteristics with IRAS 16547$-$4247, the object observed by Garay et 
al. (2003). The source \object{IRAS~16353$-$4636} has been cataloged as 
a star-forming region by Avedisova (2002), based on its IRAS colors, 
and studied for the first time by Combi et al. (2004). Although the 
available multi-wavelength data from radio to high-energy gamma rays 
suggested it could be a microquasar candidate, later X-ray observations 
with {\it XMM-Newton} provide a more precise position that was away 
from the IRAS source and its radio counterpart (Bodaghee et al. 2006). 
In fact, the X-ray source has been found to be an accreting X-ray 
pulsar.

The main goal of this study is to establish the nature of 
\object{IRAS~16353$-$4636}. Section 2 describes the already known sources 
in the target field, and details the observation and reduction 
processes we have carried out. In Sect. 3 we present the 
results. A discussion is given in Sect. 4 and the conclusions are 
stated in Sect. 5.

\section{Observations and data reduction} \label{observations}

\subsection{The field of IRAS~16353$-$4636}

The infrared source is centered at ($\alpha, \delta$) [J2000] = $16^{\rm 
h} 39^{\rm m}$ $3.52^{\rm s}, -46^\circ 42' 28.28''$. The uncertainty ellipse
major and minor axes and position angle are $14'' \times 4''$, 97$^\circ$ 
(Beichman et al. 1988). Mid-IR and Far-IR fluxes are given in 
Table 1. Sources from other catalogs are found in the neighborhood and 
we plotted them in Fig. 1. The MSX6C catalog (Egan et al. 2003) 
lists the source G337.9947$+$00.0770, which coincides with 
IRAS~16353$-$4636 
and is detected at the four MSX bands (Table 1).

In the frame of the RMS Survey, Urquhart et al. (2007) have detected 
$^{13}$CO (J=1--2) emission lines from the MSX source, using the Mopra 
Telescope (angular resolution: 20$''$, velocity resolution: 0.2 km 
s$^{-1}$, and noise temperature: 0.1 K). Very recently, Mottram et al. 
(2010) have estimated the MIPSGal 70$\mu$m flux of the RMS source (Table 
1).

In addition, there are two Spitzer GLIMPSE point sources (SSTGLMC 
G337.9986$+$0.0758 and G337.9907$+$0.0733, Benjamin et al. 2003, Churchwell 
et al. 2009) in the vicinity of our target, though neither of them is 
positionally coincident with IRAS 16353$-$4636.

The X-ray source IGR J16393$-$4643 (Bodaghee et al. 2006, Chaty et al. 
2008, Corbet et al. 2010), which is of X-ray pulsar origin, is also 
plotted in Fig. 1. The region also hosts a low-frequency radio source, 
J163903.9$-$464215.55, detected with the Giant Millimetre Radio 
Telescope (GMRT; Pandey et al. 2006) at 610 MHz. There is no spatial 
correlation between IRAS~16353$-$4636 and the X-ray and GMRT-radio 
sources.

\begin{table}[h]
\begin{center}
\caption{Relevant infrared sources in the field.}
\label{table:ired}
\begin{tabular}{l@{~~~}l@{~~~}r@{~~~}r@{~~~}r@{~~~}r}
\hline
\hline
Source & $\alpha,\delta$[J2000] & $\lambda$      & Flux & Size / \\
       &  (hms,dms)             & ($\mu$m) & (Jy) & ang.res \\ 
\hline
IRAS & 16 39 03.5, & ~12 & 13.5$\pm$0.7 & $14''\times 4''$\\ 
16353$-$4636    &  $-$46 42 28   &  ~25 & 80.5$\pm$4.0~ &  
P.A.=97$^\circ$\\ 
              &                &  ~60 &  $<$806 & \\
              &                & 100 &  2210$\pm$330 &  \\
MSX   &  16 39 03.4, & 8.28  &  6.339$\pm$4.1\% &$18'' \times 18''$ \\
G337.9947  &  $-$46 42 27 &  12.13  & 9.454$\pm$5\%   & \\
$\,\,+0.0770   $              &     &  14.65  & 8.788$\pm6.1\%$ &  \\
     &             &  21.34  & 35.59$\pm$6\%  &  & \\
MIPSGal$^\dag$ & 16 39 03.4,    &  ~70 & 698.24$\pm$5.24 &$2'' \times 2''$\\
               &  $-$46 42 27  &     & \\       
\hline
\end{tabular}
\end{center}
$\dag$: Flux value from Mottram et al. 2010.
\end{table}

\begin{table}[ht]
\begin{center}
\caption{ATCA observational parameters, and integrated flux density at each 
frequency.}
\label{table:atca}
\begin{tabular}{r@{~~~}c@{~~~}c@{~~~}c@{~~~}l@{~~~}c@{~~~}r}
\hline
\hline
$\nu$ & Conf. & $t_{\rm int}$ & Weight$^{\dag\dag}$ & Synth. beam, 
& rms & $S_{\rm int}$\\
(GHz) &   & (min)  &   &    Position angle & (mJy/b)  & (mJy) \\
\hline
 1.384 & 6A & 247 & $R$=0  & 7\farcs7$\times$5\farcs9, $\,\,\,-$3.3\degr\ & 1.00 &90$\pm4$ \\
 2.368 & 6A & 247 & $R$=0  & 4\farcs6$\times$3\farcs7, $\,\,\,-$4.3\degr\ & 0.50 & 2$\pm$4 \\
4.800 & 6A & 201 & $R$=0  & 2\farcs4$\times$1\farcs6, $-$16.6\degr\ & 0.14 & 72$\pm$3 \\
 8.640 & 6A & 201 & $R$=0  & 1\farcs4$\times$0\farcs9, $-$19.2\degr\ & 0.09 & 31$\pm$5 \\
 17.344& 6C & 528 & $R$=2 & 1\farcs0$\times$0\farcs7, $\,\,\,+$1.4\degr\ & 0.10 & 81$\pm$20 \\
 19.648& 6C & 528 & $R$=2 & 1\farcs0$\times$0\farcs6, $\,\,\,-$0.2\degr\ & 0.09 & 65$\pm$15 \\
\hline
\end{tabular}
\end{center}
Note: mJy/b: mJy beam$^{-1}$. $\dag\dag$: R = Robust, see text.
\end{table}

\begin{figure} 
\begin{center}
\includegraphics[width=7cm,angle=-90]{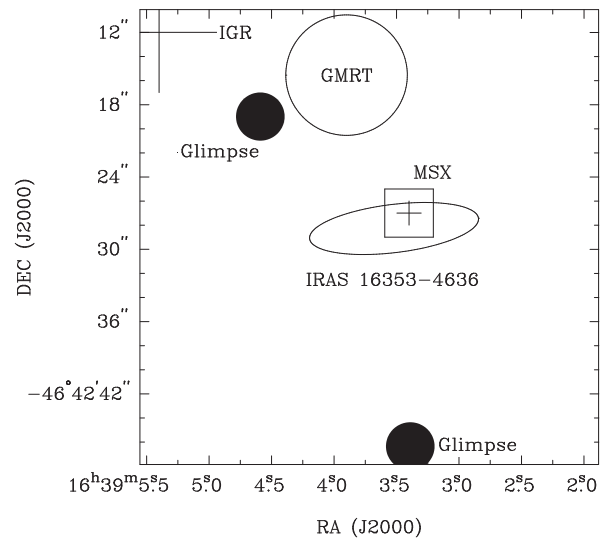}
\caption{IRAS~16353$-$4636 (ellipse) and other sources in its field. 
Spitzer-GLIMPSE point sources: filled circles (Benjamin et al. 2003); 
MSX6C G337.9947$+$00.0770: square, GMRT-610 MHz J163903.9$-$464215.55 
(Pandey et al. 2006): hollow circle; IGR J16393$-$4643 (Bodaghee et al. 
2006): larger cross. Smaller cross: observations of $^{13}$CO line 
(Urquhart et al. 2007).}
\label{thefield}
\end{center}
\end{figure}

\subsection{ATCA} \label{atcaobs}

\begin{figure*}[ht] 
\begin{center}
\includegraphics[width=13cm]{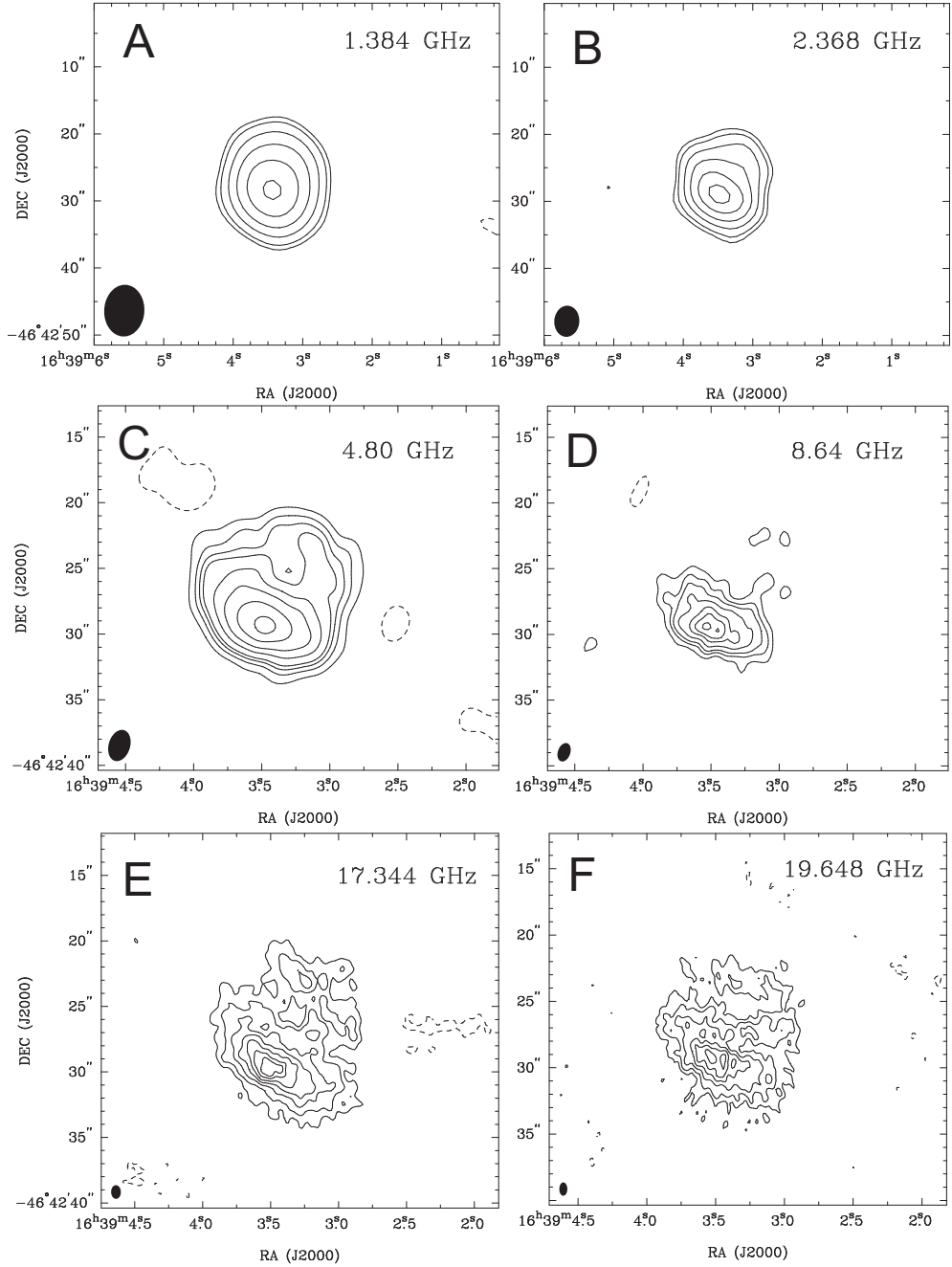}
\caption{ATCA radio continuum images of
\object{IRAS~16353$-$4636}. {\bf A}: Image at 1.4 GHz; contour 
levels of $-$3, 3, 5,
10, 18, 30, and 45 times the rms noise of 1.0~mJy~beam$^{-1}$. {\bf B}: 
Image at 2.4 GHz; contour levels of $-$3, 3, 5, 10, 18,
30, and 45 times the rms noise of 0.5~mJy~beam$^{-1}$. {\bf C}: 
Image at 4.8 GHz; contour levels of $-$3, 3, 6, 9, 12, 20, 30, 45 
and 60 times the rms noise of 0.14~mJy~beam$^{-1}$. {\bf D}:
Image at 8.64 GHz; contour levels of $-$3, 3, 6, 9, 12, 20, 24 and 28
times the rms noise of 0.09~mJy~beam$^{-1}$. {\bf E}:
Image at 17.344 GHz; contour levels of $-$3, 3, 5, 8, 11, 14, 17, 
19, and 20 times the rms
noise of 0.10~mJy~beam$^{-1}$. {\bf F}: Image at 19.648
GHz; contour levels of $-$3, 3, 5, 8, 11, 14, 17, 19, and 20 times
the rms noise of 0.09~mJy~beam$^{-1}$. Synthesized beams are 
plotted at bottom left corners. North is up and east is to the left.}
\label{fig:13and20cm}
\end{center}
\end{figure*}

We have carried out radio continuum observations toward the source 
\object{IRAS~16353$-$4636}, with the Australia Telescope Compact Array
(ATCA), at the frequencies 1.384, 2.368, 4.800, 8.640, 17.344, and 
19.648~GHz.

\paragraph{Observations in the cm range.} The data were taken in 2004 
January 11, in the 6A configuration, at full synthesis (16:00--4:00~UT). 
The field was observed interleaving simultaneous observations at two 
frequency pairs: 1.384/2.368~GHz and 4.800/8.640~GHz. 
\object{PKS~1934$-$638} served as the flux calibrator, and the nearby 
source \object{1646$-$50} was observed for phase calibration. The total 
bandwidth at these four frequencies was 128~MHz over 32 channels. The 
total time on source at each frequency resulted in 247~min at 1.384 and 
2.368~GHz, and 201~min at 4.800 and 8.640~GHz.

\paragraph{Observations in the mm range.} We observed simultaneously at 
17.344 and 19.648~GHz in 2006 March 29, using ATCA in the 6C 
configuration (13:00--0.30~UT). The sources \object{PKS~1934$-$638}, 
\object{1253$-$055} and \object{1646$-$50} were used as flux, bandpass, 
and phase calibrators, respectively. The total bandwidth was 128~MHz over 
32 channels. The total time on source was 528~min. The high phase 
stability allowed us to perform 12-min target source scans.\\

All six data sets were reduced and analyzed with the {\sc miriad} 
package. The calibrated visibilities were Fourier-transformed using 
`natural', `uniform', and `robust' weightings. The robust 
weighted images combine the `natural' lower noise 
and the `uniform' higher angular resolution (Briggs 1995). 
The robustness parameter was set to $R=0$ for the 6A (cm) data, and to  
$+2$ for the 6C (mm) data\footnote{`Robust' weights are a function of 
local $u-v$ weight density: in regions where the weight is low [high], 
the effective weighting is natural [uniform]. The optimal value thus 
depends strongly on the configuration. Larger values of $R$ can produce 
beams that have nearly the same point source sensitivity as the 
naturally weighted beam, but with enhanced resolution.}. 
The resulting synthesized beams and rms are given in 
Table~\ref{table:atca}.

\subsection{VLA data} \label{vlaobs}

We searched for data in the VLA archives and found observations centered 
relatively close to IRAS 16353$-$4636, taken in 2001 February 1 at 20~cm 
in the BnA configuration. The data were taken with one IF centered at the 
rest frequency of the hyperfine transition of \ion{H}{i} (1420.406~MHz), 
with 127 channels over a total bandwidth of 1.56~GHz, giving a width of 
12.2~kHz (or 2.58~km~s$^{-1}$) per channel. After Hanning smoothing, this 
resulted in a velocity resolution of 5.16~km~s$^{-1}$ per channel. The 
spectral observations were centered at $v_{\rm LSR} = 0$~km~s$^{-1}$. The 
data were edited, calibrated, and imaged using the software package 
Astronomical Image Processing System (AIPS) of NRAO\footnote{The National 
Radio Astronomy Observatory is a facility of the National Science 
Foundation operated under cooperative agreement by Associated 
Universities, Inc.}. The resulting synthesized beam was $10\arcsec\times 
3\farcs8$.

\subsection{NTT NIR} \label{nttnirobs}

We conducted photometric ($J$, $H$ and $K_{\rm s}$ filters) and spectroscopic
(0.9--2.5~$\mu$m) NIR observations of \object{IRAS~16353$-$4636} on 2004 July
10 with the spectro-imager SofI, installed on the ESO New Technology Telescope
(NTT). The large field imaging of SofI's detector was used, giving an 
image scale of 0\farcs288 pixel$^{-1}$ and a field of view of
4\farcm94$\times$4\farcm94. 

We repeated a set of photometric observations for each filter with 9 
different 30\arcsec\ offset positions including 
\object{IRAS~16353$-$4636}, with an integration time of 90 seconds for 
each exposure, following the standard jitter procedure that allows 
a clean subtraction of the blank sky emission in NIR.

The IRAF (Image Reduction and Analysis Facility package) suite was used 
to perform the data reduction, which included flat-fielding and NIR sky 
subtraction. For the three images, one in each filter, we 
obtained an astrometric solution by using more than 400 coincident 2MASS 
objects, with a final rms of 0\farcs07 in each coordinate.

We carried out aperture photometry and transformed the instrumental 
magnitudes into apparent magnitudes with the standard relation: ${\rm 
mag_{app} = mag_{inst} - Zp -~}{ext\times AM}$, where $\rm mag_{app}$ and 
$\rm mag_{inst}$ are respectively the apparent and instrumental 
magnitudes, Zp is the zero-point, $ext$ the extinction, and $AM$ the 
airmass. The observations were performed through an airmass close to 1.

The spectroscopic observation consisted of 12 spectra with the Blue and 
Red grisms of 1000 resolution, and a wavelength coverage of 9000 to 
25000 $\AA$. The position of \object{IRAS~16353$-$4636} in the slit was 
offset 30\arcsec\ in half of the exposures to subtract the blank NIR 
sky. The total integration time was 240~s in each grism. We took Xe lamp 
exposures to perform the wavelength calibration. The $5'$-long slit 
was centered at one of the most intense NTT sources that overlaps the 
radio emission ($\alpha,\delta$[J2000] = $16^{\rm h} 39^{\rm m} 
3.18^{\rm s}, -46^\circ 42' 31.48''$). The NTT source was labeled \#5 
(see Sect. 3.3 and Fig. 5). The position angle of the slit was 
$+15^{\circ}$ (positive from north to east), and its width, $1"$.

The NIR spectra were reduced with IRAF by flat-fielding, correcting 
the geometrical distortion using the arc frame, shifting the individual 
images using the jitter offsets, combining these images, and finally 
extracting the spectra. The analysis of SofI spectroscopic data, and more 
precisely the sky subtraction, was difficult due to a variable sky, mainly 
in the red part of the blue grism, causing some wave patterns. We observed 
a photometric standard star Hip084636 (G3 V) from the Catalog of Persson et 
al. (1998), to correct for telluric absorption. The spectra
were finally shifted to the {\sc lsr} rest frame.

\section{Results} \label{results}

\subsection{ATCA} \label{atcares}

The continuum images built with the ATCA data are shown in Fig.~2. The 
angular resolutions range from $\sim 8\arcsec$, at the lowest radio 
frequency, to less than 1\arcsec\, at the highest one. The radio source 
is detected at all frequencies. The position of the peak flux (PFP) at 
all frequencies agrees among position uncertainties. At the highest 
angular resolution (19.6 GHz) we measure $(\alpha, \delta)_{\rm 
PFP}[{\rm J}2000] = 16^{\rm h} 39^{\rm m} 3.53^{\rm s}, -46^\circ 42' 
29.3''$. The integrated flux density $S_{\rm int}$ is derived summing 
all the flux above the 3$\sigma$ contour (1$\sigma$ = image rms). We 
adopted an error in the integrated flux density equal to the absolute 
difference between the integrated flux above 1$\sigma$ and $S_{\rm int}$. 
Table 2 lists the results. Throughout the text we have followed the 
convention $S_{\nu} \propto \nu^{\alpha}$, i.e. a spectral index $\alpha 
> 0$ for thermal emission, and $\alpha < 0$ for non-thermal emission.

At the lowest frequency --1.4 GHz, and the largest beam-- no structure is 
appreciated. At 2.4 GHz extended emission is detected toward the N-W, though 
it is better defined at 4.8 GHz. The detections from 4.8 to 19.6 GHz 
show rather elongated sources, in the direction SW-NE. The images at 17.3 
and 19.6 GHz reveal a clumpy structure.

The estimate of the spectral index can help to characterize the 
radiation regime and, eventually, can lead to a firm identification of 
the astrophysical source. We build a spectral index map at the highest 
angular resolution possible, i.e., using the data at 17.3 and 19.6 Ghz. 
To attain the same angular resolution at the two frequencies we restore 
the images to the same Gaussian beam size ($1''\times 1''$) for 
both data sets.

We consider only pixels with signal-to-noise $\geq$ 5. In Fig. 3 we 
present the resultant spectral index map. It displays fine-scale 
structure on size scales comparable to the NTT infrared sources, as 
shown by the $K_{\rm s}$ contours superimposed on to the spectral 
index map in Fig. 3.

In general, the index is negative (non-thermal) towards the radio 
peak seen in Fig. 2, and towards the south of the infrared cluster 
associated with IRAS 16353$-$4636, and is positive (thermal) towards 
the north and west.

Radio observations from massive YSOs have revealed both thermal (as in 
Mart\'{\i} et al. 1995) and non-thermal outflows or jets (like in 
Rodr\'{\i}guez et al. 1989). Strong shocks can occur at the end points 
of the jets, giving rise to diffusive shock particle acceleration 
which, in turn, will produce non-thermal emission of synchrotron origin 
(see, for instance, Romero 2010, and references therein).

The rough matching of the 17.3/19.6 GHz spectral index distribution with the 
NTT image in Fig. 3 shows that {\sl (i)} the strongest of the southern sources 
(here called \#5, see below) is superimposed on a radio emitting region with 
$\alpha << 0$; {\sl (ii)} at the position of the strongest of the 
eastern sources (\#8, see below), $\alpha > 0$; and {\sl (iii)} a negative 
spectral index corresponds to the {\sc pfp} radio maximum.

\begin{figure} 
\begin{center}
\includegraphics[width=6.5cm,angle=-90]{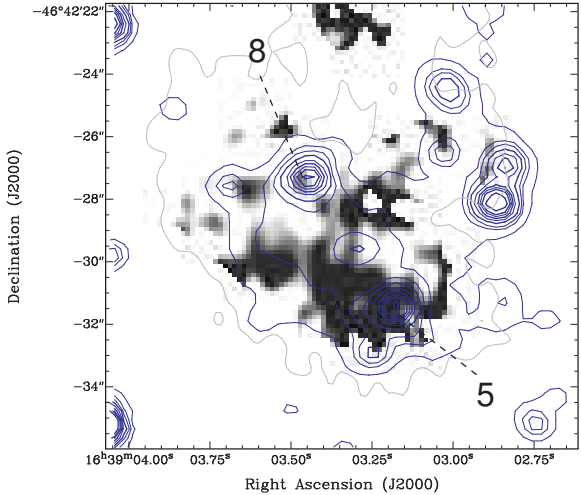}
\caption{Spectral index map derived for the
17.344~GHz/19.648~GHz ATCA data, in grayscale: black is $-$1, white is 
$+$1.
The NTT $K_{\rm s}$-band emission is superimposed as blue 
contours of 25, 50, 100, 150, 200, 300, and 400 Jy.
The grey contour corresponds to 3$\sigma$ level 
continuum emission at 17.344 GHz. NTT sources here named 5 and 8 are 
marked (see text).}
\label{fig:spixs17-19}
\end{center}
\end{figure}


\subsection{VLA} \label{vlares}

We produce a continuum image at 1.42 GHz using the line-free 
channels in the $+$5 to $+$152~km~s$^{-1}$ range. It basically 
matches the ATCA image obtained at 1.384 GHz, thus confirming  
the results of the ATCA observations albeit with pointing 
differences. These are probably due to poor phase referencing at 
low elevations for the VLA.

\begin{figure}[!ht] 
\center
\includegraphics[width=7.5cm]{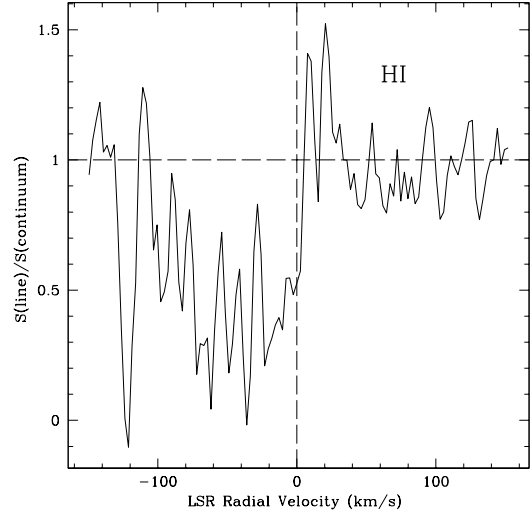}
\caption{21-cm \ion{H}{i} absorption spectrum for the target, 
plotted as a
function of {\sc lsr} radial velocity. The measured spectrum is divided  
by the
continuum level and represents $e^{-\tau}$, where $\tau$ is the optical depth. 
The horizontal dashed line is
drawn at $S_{\rm line}/S_{\rm continuum}=1$ and the vertical dashed 
line is drawn at an {\sc lsr} radial velocity of 0~km~s$^{-1}$.}
\label{fig:vla-hi}
\end{figure}

From the VLA data set, we also obtain \ion{H}{i} spectra toward 
the peak in these radio data. The center of the box region over 
which the \ion{H}{i} emission was averaged coincides with star \#5 in 
our near-IR observations (see below). However, due to the larger angular 
resolution of the VLA data, the region encompassed all the near-IR sources 
relevant here. The \ion{H}{i} spectrum of the target is presented in 
Fig.~\ref{fig:vla-hi}, Hanning-smoothed to a resolution of
5.2 km s$^{-1}$. At 
negative velocities there is broad absorption down to $-100$ km 
s$^{-1}$, followed by a detached feature spread over 4 channels,
centered at a velocity of approximately $-$120 km s$^{-1}$.

\begin{figure*} 
\begin{center}
\includegraphics[angle=0,width=12cm]{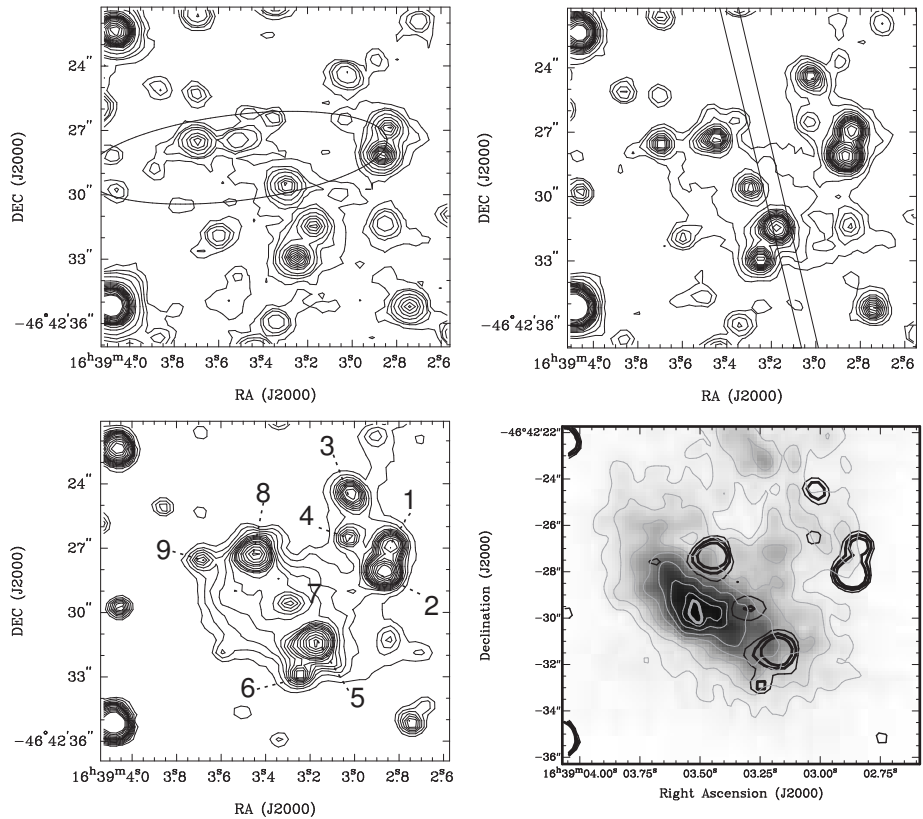}
\caption{{\sl Upper left panel:} $J$-band image of the region toward
\object{IRAS~16353$-$4636},
obtained with the NTT on 2004 July 10. Levels: 1, 3, 5, 10, 20, ..., 
100, 120, 150, 190, 250, 320, and 400 Jy. 
The uncertainty in the position of
the IRAS source is indicated by an ellipse.
{\sl Upper right panel:} The same as the upper left panel
but in the $H$ band. Levels: 5, 10, 20, ..., 400 Jy.
We show the position and size of the slit
used for spectroscopic purposes.
{\sl Lower left panel:} The same as the upper left panel but in the 
$K_{\rm 
s}$
band. Levels: 20, 30, ..., 400 Jy. The extended
emission appears more clearly than in the $J$- and $H$-band images.
Several point-like sources are visible, and labeled with numbers as in 
Table 3, at increasing right ascension.
{\sl Lower right panel:}
Continuum emission at 17.344 GHz in greyscale and grey 
contours, superimposed on to black contours corresponding to strongest
near-IR $K_{\rm s}$ sources.
North is up and east is to the left.}
\label{nttres}
\end{center}
\end{figure*}

\subsection{NTT NIR} \label{nttnirres}

We show the final $J$, $H$ and $K_{\rm s}$ band images in Fig. 5; 
we also compare our 17.3 GHz radio map with the $K_{\rm s}$ 
band image in this figure.
Besides diffuse NIR emission, there are several point-like 
objects. We have identified the brightest ones, 
and numbered them from 1 to 9 at increasing right ascension
(labeled in Fig. 5, bottom left panel). 

The {\sl daophot} package in the IRAF suite has been used to extract 
all the sources in the crowded field through PSF fitting. The aperture 
has been adapted from 4 to 10 pixels, up to the Airy's ray if possible, 
to each source, only to take into account the stellar flux, subtracting 
the nebula flux. At the end of the extraction process, we apply an 
aperture correction so as to finally use the same aperture photometry 
for all of the sources. An adjacent annulus of 1 pixel outer 
radius is used to estimate the sky background. The results are quoted 
in Table~\ref{table:nir}. The magnitudes of the whole extended emission 
(point-like sources and diffuse emission) are $J=13.1\pm0.5$, 
$H=11.9\pm0.2$ and $K_{\rm s}=10.7\pm0.2$.

The NTT photometry is used to build a color-color (C--C) plot. $(J-H)$ 
versus $(H-K_{\rm s})$ colors are plotted in Fig. 6. As a reference, we 
have included the loci of dwarf, giant and supergiant stars, following 
the intrinsic colors listed by Tokunaga (2000). Straight dotted lines 
show the reddening vectors for these stars, and define a band where 
stars with normal colors, only affected by reddening, are expected to 
be found in the diagram. All the objects lying to the right of this 
band show an infrared excess in their $H-K_{\rm s}$ colors and are 
{\bf presumably young stars still undergoing some form of accretion}.

From a global point of view, all the objects in Fig. 6 seem to share a 
common amount of reddening, equivalent to 8 to 10 magnitudes of 
extinction in the visual band. The colors of objects \#1, 2, 3, and 4 
are characteristic of reddened normal stars like foreground objects. 
The rest of the sources shows an infrared excess at $K_{\rm s}$-band; 
they could represent an embedded cluster of YSOs. We have examined in 
particular the two brightest $K_{\rm s}$ sources with colors of 
protostellar objects (stars \#5 and \#8). We use NTT spectroscopy, and 
SED fitting to study \#5 and \#8, respectively. Notably, source 8 lies 
close to the IRAS and MSX peak in this region, and is the most extreme 
outlier in the C--C diagram.  Source 8 also coincides with a region 
of positive spectral index in our radio data (Fig. 3).  This is 
consistent with thermal radio emission which could be 
associated with a thermal radio jet or optically thick HII region 
(Mart\'{\i} et al. 1993). Source 8 is therefore potentially the best 
candidate for an intermediate or even high-mass YSO in the cluster. 
Source 5, on the other hand, coincides with non-thermal radio 
emission, as one might expect if this source was also a YSO associated 
with an outflow (see for example Bosch-Ramon et al. 2010 for 
theoretical considerations). It is unlikely, however, that source 5 is 
a UCHII region, since UCHII regions are usually associated with thermal 
emission (Wood \& Churchwell 1989).

Figure 5, bottom right panel, shows the correlation of near-IR 
emission ($K_{\rm s}$ band) with radio emission at 17.3 GHz. The radio 
peak has no near-IR {\bf counterpart}. The comparison tells us that the 
spectral index/NTT sources coincidences {\bf evident in Fig. 3} must be 
taken with caution, 
since the radio emission associated with NIR sources 5 and 8 appears to 
be swamped by extended emission associated with the radio peak itself. 
Sources 5 and 8 are not discretely resolved in the radio data. So its 
not entirely clear that the negative and positive spectral indices 
observed around sources 5 and 8 are really linked to these NIR point 
sources.  The bulk of the radio emission may be unrelated to the NIR 
sources.

\begin{table*} 
\begin{center}
\caption{NIR coordinates, the band at which they are measured 
($J$-band: 1.25$\mu$m, $H$-band: 1.65$\mu$m, 
$K_{\rm s}$-band: 2.17 $\mu$m), and magnitudes of the 9 objects, ordered 
by increasing right ascension and marked in Fig. 5, lower panel.}
\label{table:nir} 
\begin{tabular}{l c c c r r r}
\hline
\hline
Star \#& $\alpha$[J2000.0] & $\delta$[J2000.0]  & Band & $J$ & $H$ & $K_{\rm s}$ \\
       &   (h,m,s)          &($^\circ$,$'$,$''$) &      & & & \\
\hline
1  & 16 39 02.84  & $-$46 42 26.99 & $H$ & $15.12\pm0.04$ & $13.26\pm0.04$ & $12.41\pm0.07$ \\
2  & 16 39 02.87  & $-$46 42 28.09 & $H$ & $15.19\pm0.05$ & $13.35\pm0.05$ & $12.56\pm0.09$ \\
3  & 16 39 03.02  & $-$46 42 24.49 & $H$ & $17.11\pm0.11$ & $14.98\pm0.06$ & $13.61\pm0.08$ \\
4  & 16 39 03.01  & $-$46 42 26.58 & $K_{\rm s}$ & $16.26\pm0.07$ & $14.18\pm0.06$ & $13.37 \pm0.14 $ \\
5  & 16 39 03.18  & $-$46 42 31.47 & $H$ & $15.28\pm0.05$ & $13.78\pm0.05$ & $12.63\pm0.09$ \\
6  & 16 39 03.25  & $-$46 42 32.92 & $J$ & $15.30\pm0.05$ & $13.93\pm0.05$ & $12.64\pm0.08$ \\
7  & 16 39 03.30  & $-$46 42 29.59 & $J$ & $15.99\pm0.07$ & $14.69\pm0.10$ & $13.58\pm0.15 $ \\
8  & 16 39 03.45  & $-$46 42 27.34 & $K_{\rm s}$ & $16.42\pm0.17 $ &$14.80\pm0.21 $ & $13.04 \pm0.17 $ \\
9  & 16 39 03.70  & $-$46 42 27.54 & $J$ & $16.37\pm0.11$ & $15.00\pm0.11$ & $13.40\pm0.11 $ \\
\hline
\end{tabular}
\end{center}
Astrometric errors are: $\Delta \alpha < 0.01^{\rm s}$, $\Delta \delta < 
0.1''$.  
\end{table*}

\begin{figure} 
\begin{center}
\includegraphics[width=9cm,angle=-0]{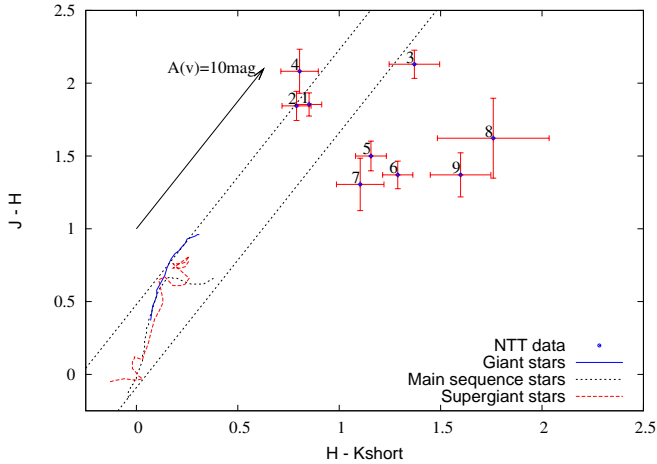}
\caption{Near-infrared color-color plot $(J-H)$ vs $(H-K_{\rm s})$ 
derived 
using the magnitudes measured by us with the NTT, listed in Table 3. 
The dotted curves 
show the loci of dwarf, giant and supergiant stars, based on the 
intrinsic colors by Tokunaga (2000). The straight dotted
lines represent the reddening vectors for these stars (see text).}           
\end{center}
\end{figure}

\subsection{Individual sources}

\noindent {\bf Source \#5.} Our NTT spectrum of source 5 reveals the 
presence of \ion{H}{i} emission lines and absorption in the CO vibrational 
bands ($^{12}$CO and $^{13}$CO, $\Delta v =2$ and 3), typical of pre-main 
sequence stars (see Nisini et al. 2005 and references therein). Figure 7 
shows the spectrum with individual lines identified in the $J$, $H$, and 
$K_{\rm s}$ bands\footnote{We checked the lines at {\tt 
http://www.jach.hawaii.edu/\\UKIRT/astronomy/calib/spec$\_$cal/lines.html}}.

We find the following spectral lines: Pa$\beta$, Pa$\delta$, $^{12}$CO 
lines and bands, $^{13}$CO bands, Br$\gamma$, and possibly Br$\,10$, 
Br$\,11$. All of them are characteristic of star-forming regions.  
Many lines peak at velocities close to or more negative than the 
\ion{H}{i} absorption feature at $-$120 km s$^{-1}$. Table 4  
lists the suggested line identifications, measured wavelengths, rest 
wavelengths, and measured {\sc lsr} central velocities. A standard 
Galactic rotation circular model, such as Brand \& Blitz (1993), 
assigns a minimum velocity of about $-$125 km s$^{-1}$ to the gas at 
the tangent point. {\bf Table 4 indicates that the velocity}
shifts are extremely widespread.

Errors in the determination of radial velocities are strongly dependent 
on the signal of the individual lines measured and the wavelength 
resolution of the spectrum. For relatively strong emission lines, such 
as Br$\gamma$, the uncertainty in the estimation of the centroid of the 
Gaussian profile is expected to be about 1/10 of a pixel, which is 
equivalent to about 14 km s$^{-1}$. The uncertainty grows for less 
intense profiles and largely increases for the CO lines, for which no 
profile fitting is possible as we are dealing with unresolved 
absorption bands.

Besides identifying hydrogen emission and CO absorption features that 
are typical of pre-main-sequence (PMS) interaction, we were also able 
to identify a few absorption lines, which can be attributed to the PMS 
star itself. As shown in Fig. 7, Mg~I (1.71 $\mu$m) and Na~I (2.20 
$\mu$m) seem to be present in our NIR spectrum, although somewhat 
noisy, with equivalent widths of $\sim 3 \AA$ and $\geq 2 \AA$, 
respectively. Notwithstanding the presence of spectral veiling from the 
circumstellar disk, we can perform a first-order approximation of the 
stellar spectral type, making use of the atlas of Rayner et al. 
(2009) 
and its discussion in Bik et al. (2010). From the cool star atlas, we 
find that the present combination of equivalent widths can only be 
found in a late (K5-M0) dwarf star.\\

\begin{table}[h]
\begin{center}
\caption{
Near-IR spectral lines detected in  our NTT spectrum of source \#5. 
Measured and laboratory wavelengths are listed in angstroms in cols.
2 \& 3, and their measured radial velocities ($v_{\rm 
LSR}$), are shown in col. 4. 
Radial velocities derived for CO lines are indicated in brackets as
our spectral resolution does not allow us to identify individual lines
that constitute the absorption band. We have therefore manually
measured the wavelength of the edge of the absorption profile and used
it as a reference value for the band head.}
\label{table:nttcolines}
\begin{tabular}{l l c r r}
\hline
\hline
Line ID & $\lambda$ & $\lambda_0$ & $v_{\rm LSR}$ \\
        & ($\AA$)  &  ($\AA$)   &  (km s$^{-1}$)\\
\hline
Pa$\,\delta$ &  10045.0 & 10052 & $-$220\\
Pa$\,\beta$  & 12814.5  & 12822 & $-$186\\
$^{12}$CO (3 -- 0)  & 15578  & 15582    & ($-$88) \\
$^{12}$CO (4 -- 1)  & 15776  & 15780 &  ($-$87)\\
$^{12}$CO (5 -- 2)  & 15978.2  & 15982 & ($-$82)\\
$^{12}$CO (8 -- 5)  & 16609.2  & 16610  & ($-$29)\\
Br$\,$11--4   &  16800.5: & 16811 & $-$198\\
Mg I          &  17113.3  & 171200:  & --\\
Br$\,$10--4   &  17369.3 &  17367 & $-$127\\
Br$\,\gamma$  & 21651.4  & 21661  & $-$137\\
Na I          & 22050:  & 22062  & -- \\
Ca I          & 22620:  & 22614  & -- \\  
$^{12}$CO (2 -- 0) & 22928.6 & 22935 & ($-$95)\\
$^{12}$CO (3 -- 1) & 23222.4 & 23227 & ($-$71)\\
$^{13}$CO (2 -- 0) & 23448 & 23439.1  & ($-$125)\\
$^{12}$CO (4 -- 2) & 23516.8 & 23535 & ($-$243)\\
$^{13}$CO (3 -- 1) & 23728.2 & 23739 & ($-$148)\\
$^{12}$CO (5 -- 3) & 23819 & 23829 & ($-$156)\\
$^{12}$CO (6 -- 4) & 24130.3 & 24142 & ($-$156)\\
\hline
\end{tabular}
\end{center}
\end{table}


\begin{figure} 
\begin{center}
\includegraphics[width=6cm,angle=-90]{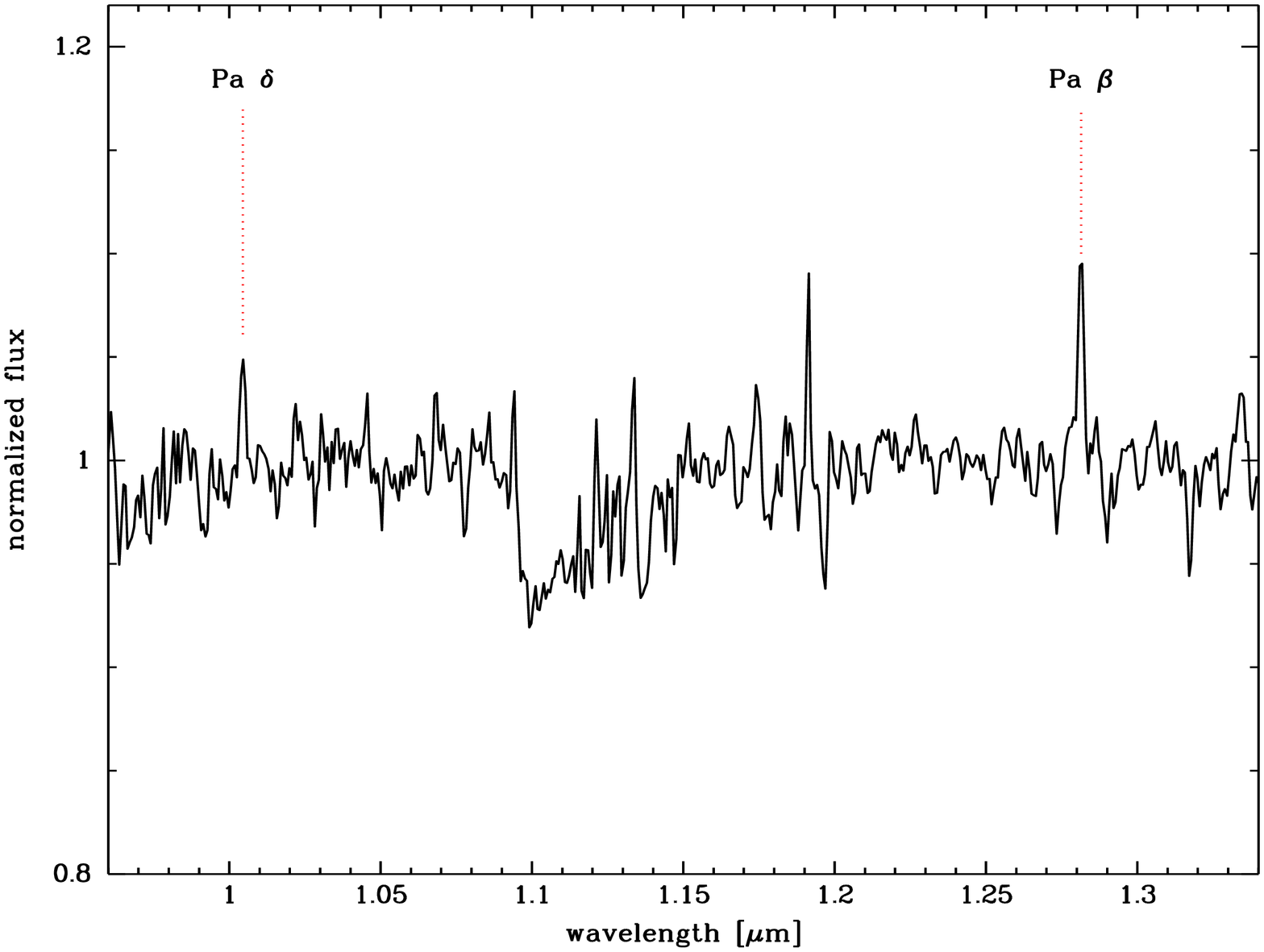}
\includegraphics[width=6cm,angle=-90]{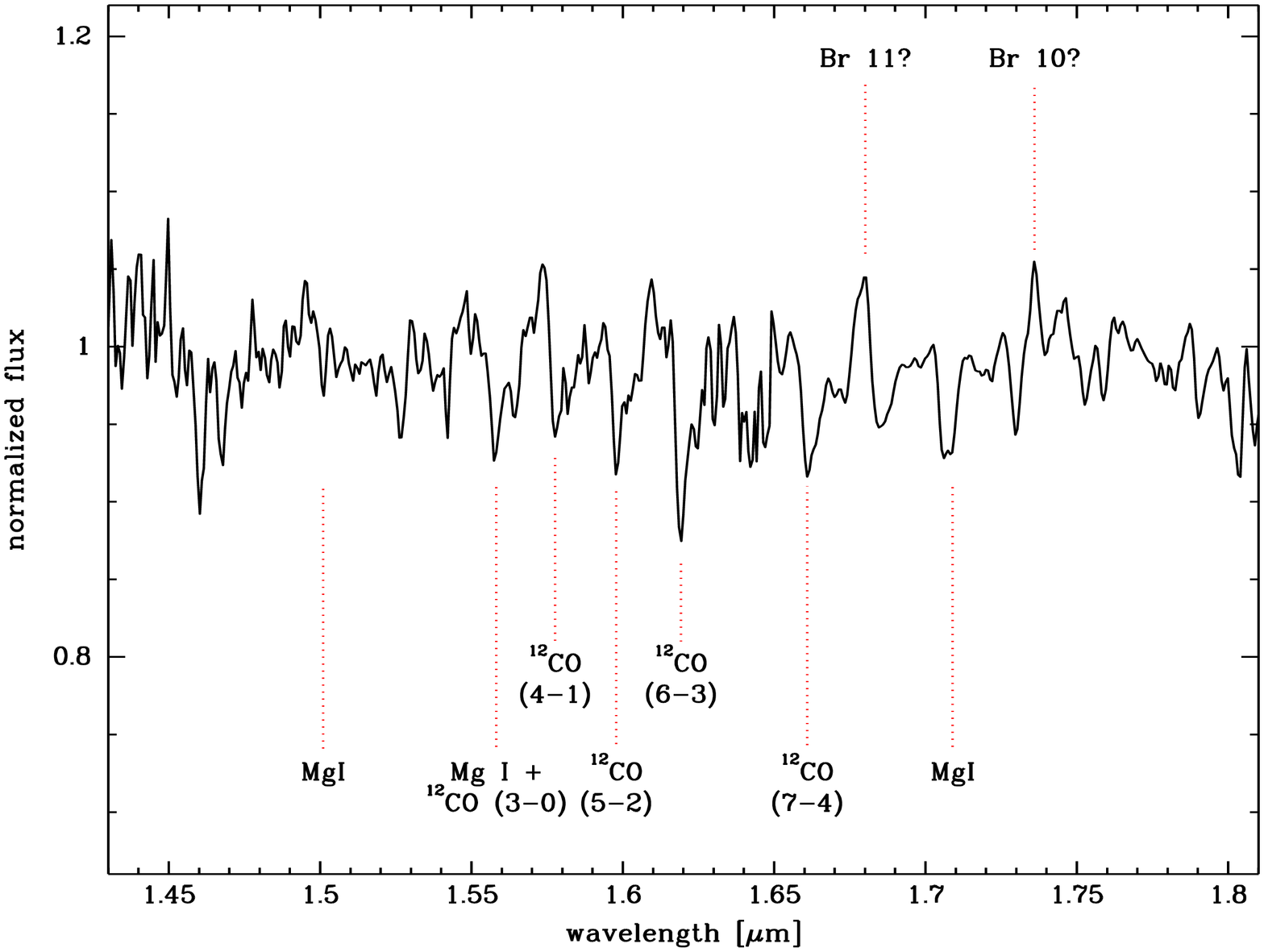}
\includegraphics[width=6cm,angle=-90]{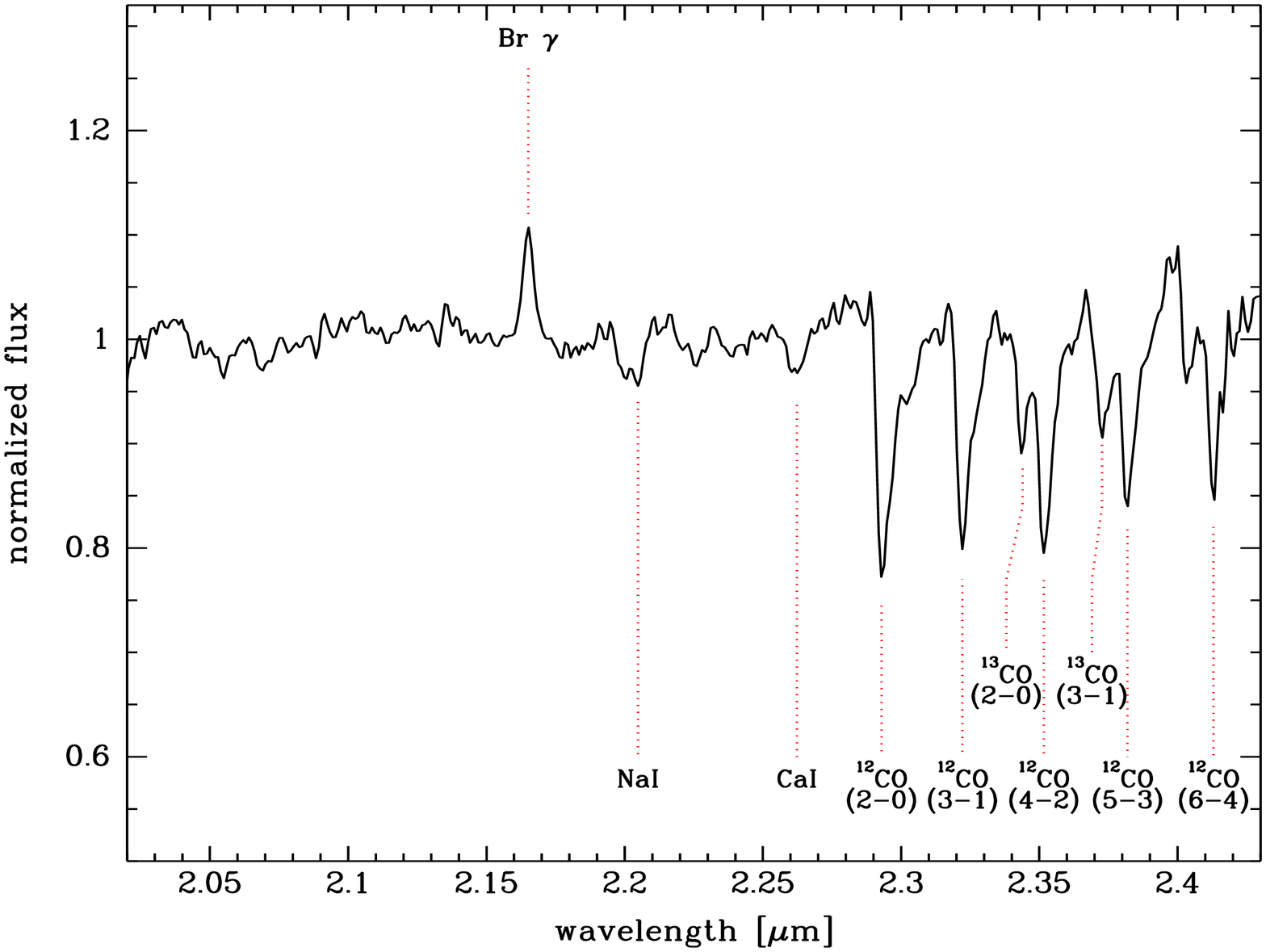}
\caption{Near-infrared spectrum of the central region of
\object{IRAS~16353$-$4636}, acquired with the NTT on 2004 July 10. The 
identified emission lines are indicated, most notably Pa$\beta$, 
Pa$\gamma$, $^{12}$CO lines and bands, $^{13}$CO bands, 
Br$\gamma$, and possibly Br$\,10$, and Br$\,11$.
{\sl Upper panel:} NIR $J$-band spectrum.
{\sl Central panel:} NIR $H$-band spectrum.
{\sl Lower panel:} NIR $K_{\rm s}$-band spectrum.}
\end{center}
\label{espectros-nir}
\end{figure}

\noindent {\bf Source \#8}. The position of the MSX source 
G337.9947+0.0770 (Sect. 2) corresponds to NTT source \#8. Mottram and 
co-workers (2010) have measured the 70-$\mu$m flux toward this MSX 
source, from the Spitzer-MIPSGal data, also noting that IRAS 60 and 
100-$\mu$m fluxes can be considered upper limit fluxes of the same 
source. We use these {\bf mid-infrared fluxes, together with our 
near-infrared photometry,} as inputs for the SED 
builder and 
fitting tool by Robitaille et al. (2007)\footnote{Available on-line at 
http://caravan.astro.wisc.edu/protostars/}. The authors have built a 
grid of precomputed radiative transfer models that use a fast $\chi^2$ 
minimization algorithm. If $N$ is the number of data points (excluding 
the upper limits), we select the two models for which $\chi^2 - 
\chi_{\rm min}^2 < 2N$. In Figure 8 we plot the best model that 
fulfills the last condition. The SED has been built with the NTT 
$JHK_{\rm s}$ fluxes, the MSX fluxes, the 70$\mu$m-Spitzer-MIPS flux, 
and the IRAS 60 and 100$\mu$m fluxes as upper limits. We have set a 
distance range from 6 to 10 kpc (see Sect. 4), and an interstellar 
visual absorption range from 5 to 12 (Sect. 3.3).

\begin{figure} 
\begin{center}
\includegraphics[width=7cm,angle=0]{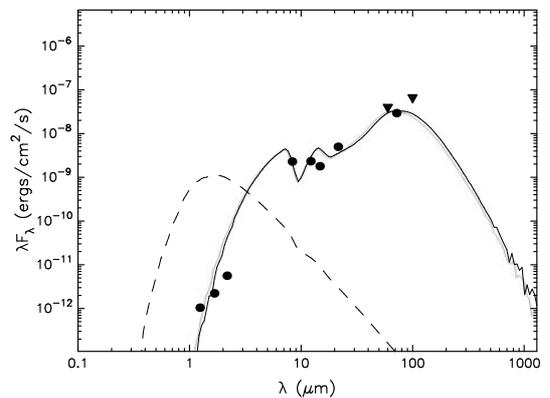}
\caption{SED fit, using the web tool as explained in Robitaille et al. (2007).
The dots represent flux values from NTT source \#8 (this work), MSX6C, 
70$\mu$m Spitzer-MIPSGal, and 60+100$\mu$m IRAS data.}
\end{center}
\end{figure}

\section{Discussion} \label{discussion}

The \ion{H}{i}-line self-absorption technique is a useful tool 
when deriving distances (see, for instance, Busfield et al. 2006). In 
this way,
we measure the most negative absorption feature of the VLA-data 
spectrum on IRAS 16353$-$4636 (Fig. 4) at $v_{\rm neg} \sim -120$ km 
s$^{-1}$. 
The Galactic rotation model of Brand \& Blitz (1993) provides two 
kinematical distances: 6.8 and 9 kpc. 
The presence of absorption at $-$120~km~s$^{-1}$ indicates that the
source is located at or beyond the sub-central point at a distance of
$\sim$8~kpc. At such a distance, 1\arcsec\ corresponds to 0.04~pc, and
the $\sim$10\arcsec\ size of the radio source, to 0.4~pc. These are
typical values for the size of a molecular cloud in the process of
contraction to form stars (e.g. Garay et al. 2003). The lack of \ion{H}{i}
absorption at positive velocities implies that the
source is not beyond the solar circle, which sets an upper limit of
$\leq$15 kpc to its distance.

The $^{13}$CO line results from Urquhart et al. (2007) confirm that 
there is molecular gas with $v = v_{\rm neg}$, at least related to NTT 
source \#8. The Mopra profile shows five emission lines, with {\sc 
lsr}-velocities from $-122.7$ to $-38.6$ km s$^{-1}$. Table 5 lists the 
components, widths, and kinematic distances derived from them (Urquhart 
et al. 2007). The central velocity of the strongest line is in very 
good agreement with the extreme \ion{H}{i} absorption feature at 
$-$120 km s$^{-1}$. The rest of the detected $^{13}$CO lines can be 
explained, for example, if additional gas clouds are present in the 
direction of IRAS 16353$-$4636. Observations with better angular 
resolution can help to clarify this issue.

The velocities measured from the near-IR emission and absorption 
lines in Table 4 are generally consistent with the large negative 
velocities measured in $^{13}$CO.  Note in particular that the strongest 
emission line, Br$\gamma$, peaks at an {\sc lsr} velocity of $-$137 km 
s$^{-1}$ (i.e. within 15 km s$^{-1}$ the most negative $^{13}$CO 
component).

\begin{table}[h]
\begin{center}
\caption{Parameters derived from the $^{13}$CO profile pointing at source 
\#8 ($20''\times 20''$ angular resolution, Urquhart et al. 2007).}
\label{table:13colines}
\begin{tabular}{r@{~~~}c@{~~~}r@{~~~}r@{~~~}r}
\hline
\hline
$v_{\rm LSR}$ &{\sl{\sc fwhm}} & $T\,dv$ &  near-$d$  & far-$d$ \\
     (km/s)    &  (km/s)        & (K km/s)&   (kpc)    &   (kpc) \\ 
\hline
$-$122.7 & 3.3 & 29.6 & 6.9$\pm$0.74 & 8.9$\pm$0.76 \\
$-$70.3 & 6.3 & 10.3 & 4.6$\pm$0.44 &11.1$\pm$0.44 \\
$-$57.4 & 6.2 &  9.8 & 4.1$\pm$0.48 &11.7$\pm$0.48 \\
$-$47.3 & 6.4 & 20.1 & 3.5$\pm$0.54 &12.2$\pm$0.54 \\
$-$38.6 & 7.8 & 20.6 & 3.0$\pm$0.06 &12.8$\pm$0.60 \\
\hline
\end{tabular}
\end{center}
\end{table}

The bolometric luminosity of IRAS 16353$-$4636, derived using the 
IRAS flux values, is 

$$L_{\rm IRAS} = {\left({D} \over {8~{\rm kpc}}\right)}^2 ~6.1 \times 10^4 
~{\rm L}_\odot,$$

\noindent which is a typical value for massive YSOs (Garay et al. 
2003). 

On the other hand, if we assume that the continuum emission from NTT 
source \#8 has an optically-thin free-free component with flux in the cm 
wavelengths of the order of 10~mJy (the total flux of the entire radio 
source at 6cm is $\sim$ 80 mJy), and an electron temperature of $10^4$~K 
for the ionized gas, we find that an ionizing photon flux of $N_{\rm i} 
\sim 5 \times 10^{47}~{\rm s}^{-1}$ (Carpenter et al. 1990) is required 
to maintain the ionization. This ionizing photon flux can be provided by 
an O9.5 ZAMS star (Panagia 1973). The luminosity of such a star is $3.8 
\times 10^4\, {\rm L}_\odot$, within a factor or two of the bolometric 
luminosity estimated from the {\it IRAS} fluxes. Although the FIR colors 
could be compatible with an ultracompact \ion{H}{ii} region, the source 
does not appear in the CS survey by Bronfman et al. (1996).

The good fit of the SED model (Fig. 8) of the infrared spectrum of
source \#8, together with its location in the NIR C--C diagram 
(Fig. 6), and the agreement with the
coordinates of the IRAS and MSX sources in the region, make it the
most promising candidate for a MYSO in this cluster. The best fit for an
object at $\sim$ 8 kpc, with $A_V \sim 10$, is  
achieved for a model source of $\sim$ 23 solar masses, $\sim$90 solar 
radius, $\sim$ 9000 K effective temperature, and 5000 yr of age.

\section{Conclusions} \label{conclusions}

From the multi-frequency study presented here on IRAS 
16353$-$4636, we have found strong evidence that supports its nature as 
a star-forming region. The \ion{H}{i} absorption spectrum indicates that 
the source is located at or beyond the sub-central point at a distance 
of $\sim$8~kpc. At such a distance, the size of the radio source agrees 
with typical values for the size of a molecular cloud in the process of 
contraction to form stars. The region harbors, at least, two pre-main 
sequence stars or YSOs (NTT sources \#5 and \#8), plus three YSO 
candidates (NTT sources \#6, \#7, and \#9). Our photometry 
shows that these five point sources have infrared excess. NIR spectra 
obtained from one of these (\#5) indicate that it is a low-mass pre-main 
sequence star, whilst the combination {\bf of near- and mid-infrared} data 
suggests that another one (\#8) is a massive young stellar object.

The broad \ion{H}{i} absorption feature and multiple CO emission 
lines seen at various velocities along the line of sight towards this 
region suggest the presence of multiple cloud components.

The radio results reveal that the region is complex, with fine-scale 
structure seen at the highest frequencies. The results of a spectral 
index analysis at mm wavelengths are indicative of different physical 
conditions varying according to position. From 1.4 to 19.6 GHz, 
the radio emission presents a peak, characterized with a negative 
spectral index. This can be explained if the peak represents the 
terminal point of an outflow.

Obtaining NIR spectra for the remaining sources in this region will 
allow us to improve our knowledge of the whole star forming region, as 
high-quality spectra provide both physical insight into their sources 
and kinematical information necessary to disentangle projection effects. 
Higher angular resolution and better sensitivity observations at low 
radio frequencies to search for polarized emission, combined with 
molecular line and continuum studies, are needed to deepen our 
investigation of this interesting source.

\begin{acknowledgements}

We thank the referee, Dr. C.J. Davis, for insightful and constructive 
comments throughout the paper. We are grateful to the English 
Department at FCAG-UNLP for the English review. This research has made 
use of the NASA's Astrophysics Data System Abstract Service, and of the 
SIMBAD database, operated at CDS, Strasbourg, France. The work was 
supported by grants PICT 2007-00848, Pr\'estamo BID (ANPCyT), FCAG-UNLP 
project 11/G093, the Centre National d'Etudes Spatiales (CNES), and 
based on observations obtained through MINE: the Multi-wavelength 
INTEGRAL Network. M.R., J.A.C., and G.E.R. acknowledge support by DGI 
of the Spanish Ministerio de Educaci\'on y Ciencia (MEC) under grants 
AYA2007-68034-C03-01/2, FEDER funds and Plan Andaluz de 
Investigaci\'on, Desarrollo e Innovaci\'on (PAIDI) of Junta de 
Andaluc\'{\i}a as research group FQM322. M.R. acknowledges financial 
support from MEC and European Social Funds through a \emph{Ram\'on y 
Cajal} fellowship.

\end{acknowledgements}


\begin{thebibliography}{}

\bibitem[]{} Avedisova, V. S. 2002, Astronomy Rep., 46, 3, 193

\bibitem[]{} Beichman, C. A., Neugebauer, G., Habing, H. J., et al. 1988, IRAS Explanatory Supplement

\bibitem[]{}Benjamin, R. A., Churchwell, E., Babier, B. L., et al. 2003, 
PASP, 115, 953

\bibitem[]{} Bik, A., Puga, E., Waters, L. B. F. M., et al. 2010, ApJ, 
713, 883

\bibitem[]{} Bodaghee, A., Walter, R., Zurita Heras, J.~A., et al.
2006, A\&A, 447, 1027

\bibitem[]{} Bonnell, I. A., Bate, M. R., \& Zinnecker, H. 1998, MNRAS, 298, 93 

\bibitem[]{} Bosch-Ramon, V., Romero, G. E., Araudo, A. T., Paredes, J. M. 2010,
A\&A, 511, 8

\bibitem[]{} Brand, J., \& Blitz, L. 1993, A\&A, 275, 67

\bibitem[]{} Briggs, D.~S. 1995, BAAS, 27, 1444

\bibitem[]{} Bronfman, L., Nyman, L.-A., \& May, J. 1996, A\&AS, 115, 81

\bibitem[]{}Busfield, A. L., Purcell, C. R., Hoare, M. G., et al. 2006, 
MNRAS, 366, 1096

\bibitem[]{} Carpenter, J. M., Snell, R. L., Schloerb, F. P. 1990, ApJ, 362, 147

\bibitem[]{} Chaty, S., Rahoui, F., Foellmi, C., et al. 2008, A\&A, 484, 783 

\bibitem[]{} Churchwell, E., Babler, B. L., Meade, M. R., et al. 2009,
PASP, 121, 213
 
\bibitem[]{} Combi, J.~A., Rib\'o, M., Mirabel, I.~F., \& Sugizaki, M.
2004, A\&A, 422, 1031

\bibitem[]{} Corbet, R. H. D, Krimm, H. A., Barthelmy, S. D. 2010, ATel 
2570

\bibitem[]{} Egan, M. O., Price, S. D., Kraemer, K. E. 2003, AAS, 203, 5708

\bibitem[]{} Garay, G., Brooks, K. J., Mardones, D., \& Norris, R. P. 
2003, ApJ, 587, 739

\bibitem[]{} Hoare, M. G., Lumsden, S. L., Oudmaijer, R. D., et al. 2005, IAU Symp 227, 370

\bibitem[]{}Lumsden, S. L., Hoare, M. G., Oudmaijer, R. D., Richards, D.
2002, \mnras, 336, 621

\bibitem[]{} Mart\'{\i}, J., Rodr\'{\i}guez, L. F., Reipurth, B. 1993, ApJ, 416, 208

\bibitem[]{} Mart\'{\i}, J., Rodr\'{\i}guez, L. F., Reipurth, B. 1995, ApJ, 449, 184 

\bibitem[]{} McKee, C., \& Tan, J. C. 2002, Nature, 416, 59

\bibitem[]{} Mottram, J. C., Hoare, M. G., Lumsden, S., L., et al., A\&A, 
510, 89

\bibitem[]{} Nisini, B., Antonucci, S., Gianini, T. Lorenzetti, D. 2005, A\&A, 
429, 543

\bibitem[]{} Osorio, M., Lizano, S., \& D'Alessio, P. 1999, ApJ, 525, 808

\bibitem[]{} Panagia, N. 1973, AJ, 78, 929

\bibitem[]{} Pandey, M., Rao, A. P., Manchada, R., et al. 2006, A\&A,
453, 83

\bibitem[]{} Persson, S.~E., Murphy, D.~C., Krzeminski, W., Roth, M., \& Rieke, 
M.~J. 1998, AJ, 116, 2475

\bibitem[]{} Price, S. D., Egan, M. P., Carey, S. J., et al. 2001, AJ, 121, 2819

\bibitem[]{} Rayner, J. T., Cushing, M. C., Vacca, W. D. 2009, ApJS, 
185, 289

\bibitem[]{} Robitaille, T.P., Whitney, B. A., Indebetouw, R., et al. 2007, 
ApJS, 167, 256

\bibitem[]{} Rodr\'{\i}guez, L.~F., \& Mirabel, I.~F. 1998, A\&A, 340, L47

\bibitem[]{} Rodr\'{\i}guez, L. F., Curiel, S., Moran, J. M., et al. 1989, 
ApJ, 346, L85

\bibitem[]{} Romero, G. E. 2010, Mem.SAI, 81, 181

\bibitem[]{} Shu, F. H., Adams, F. C., \& Lizano, S. 1987, ARA\&A, 25, 23

\bibitem[]{} Shu, F. H., Najita, J., Galli, D., Ostriker, 
E., \& Lizano, S. 1993, Protostars and Planets III, ed. E. H. 
Levy \& J. I. Lunine (Tuscon: Univ. Arizona Press),

\bibitem[]{} Tokunaga, A. T. 2000, in Allen's Astrophysical Quantities, 
A. N. Cox (Ed.), p. 143

\bibitem[]{} Urquhart, J. S., Busfield, A. L., Hoare, M. G., et al. 
2007, A\&A, 474, 891

\bibitem[]{} Varricatt, W. P., Davis, C. J., Ramsay, S., \& Todd, S. 
P. 
2010, MNRAS, 404, 661
 
\bibitem[]{} Wood, D. O. S., \& Churchwell, E. 1989, ApJS, 69, 831

\end{thebibliography}
\end{document}